\RequirePackage{fix-cm}
\documentclass[smallextended]{svjour3}       
\smartqed  
\usepackage{graphicx}
\usepackage{multirow}
\usepackage{epsfig}

\journalname{}

\title{Parity proofs of the Kochen-Specker theorem based on the Lie algebra E8}

\author{Mordecai Waegell and P.K. Aravind }

\authorrunning{M.Waegell, P.K. Aravind}

\institute{M.Waegell$^{1,2}$,P.K.Aravind$^{1}$ \at
$^{1}$Physics Department, Worcester Polytechnic Institute, Worcester, MA 01609, U.S.A.
$^{2}$Institute for Quantum Studies, Chapman University, Orange, CA. \\
\email{caiw@wpi.edu,paravind@wpi.edu}}

\date{\today}

\begin{document}
\maketitle
\begin{abstract}
The 240 root vectors of the Lie algebra E8 lead to a system of 120 rays in a real 8-dimensional Hilbert space that contains a large number of parity proofs of the Kochen-Specker theorem. After introducing the rays in a triacontagonal representation due to Coxeter, we present their Kochen-Specker diagram in the form of a ``basis table'' showing all 2025 bases (i.e., sets of eight mutually orthogonal rays) formed by the rays. Only a few of the bases are actually listed, but simple rules are given, based on the symmetries of E8, for obtaining all the other bases from the ones shown. The basis table is an object of great interest because all the parity proofs of E8 can be exhibited as subsets of it. We show how the triacontagonal representation of E8 facilitates
the identification of substructures that are more easily searched for their parity proofs. We have found hundreds of different types of parity proofs, ranging from 9 bases (or contexts) at the low end to 35 bases at the high end, and involving projectors of various ranks and multiplicities. After giving an overview of the proofs we found, we present a few concrete examples of the proofs that illustrate both their generic features as well as some of their more unusual properties. In particular, we present a proof involving 34 rays and 9 bases that appears to provide the most compact proof of the KS theorem found to date in 8 dimensions.

\end{abstract}

\section{\label{sec:Intro}Introduction}

The exceptional Lie algebra E8 plays a role in a number of physical theories such as supergravity and heterotic string theory\cite{e8refs}. Here we show that its system of root vectors can be used to exhibit a large number of ``parity proofs'' of the Kochen-Specker (KS) theorem\cite{KS1967} ruling out the existence of noncontextual hidden variables theories as viable alternatives to quantum mechanics. The fact that the root vectors of E8 could be made to serve this end was pointed out, in different ways, by Lisonek et. al.\cite{lisonek1} and by Ruuge and van Oystaeyen\cite{ruuge}. However the proof in \cite{ruuge} is unrelated to parity proofs, while \cite{lisonek1}, though proving that $2^{1940}$ parity proofs exist, does not list even a single example of such a proof. The purpose of this paper is to supplement the observations in \cite{lisonek1} and \cite{ruuge} by (a) presenting a general framework (namely, the ``basis table'') within which the parity proofs can be exhibited, (b) showing how the symmetries of E8 can be exploited to simplify the search for its parity proofs, (c) providing an overview of the parity proofs found by our search and, finally, (d) presenting a few concrete examples of the proofs in order to convey some feeling for their variety and intricacy.\\

Parity proofs of the KS theorem are appealing because they take no more than simple counting to verify. What a parity proof is, and how it accomplishes its goals, are matters that will be explained later in this paper.\\

We make a few remarks about E8, to shed light on the way it is used in this paper. For us E8 is simply a collection of 240 vectors (namely, its roots) in a real 8-dimensional Euclidean space. These vectors define the vertices of a semiregular polytope discovered by Thorold Gosset\cite{coxeter} and sometimes described by the symbol $4_{21}$. The vectors have eight real coordinates that can be chosen in a variety of ways. A particularly judicious choice, for our purposes, is the ``triacontagonal representation'' of Coxeter and Shepard\cite{leonardo}. In this representation the coordinates of the vectors are chosen in such a way that if only the first two coordinates are retained (which amounts to projecting the vectors orthogonally from eight dimensions down to two), the tips of the vectors lie at the vertices of eight regular triacontagons lying on concentric circles. Such a projection is shown in Fig.10 of \cite{leonardo}. The eight rings of thirty points are easily picked out in the figure, while the dense network of lines connecting pairs of points are projections of the 6720 edges of Gosset's polytope.\\

Two slight modifications of this figure will convert it into the Kochen-Specker diagram of E8: the first is that only one member of each pair of diametrically opposite vertices should be retained, since one is concerned with rays rather than vectors in a proof of the KS theorem; and the second is that the line segments corresponding to the edges of Gosset's polytope should be replaced by new segments that connect only pairs of vertices that correspond to orthogonal rays. Rather than construct such a diagram, we later present a ``basis table'' that conveys essentially the same information in a more useful form. The basis table is simply a listing of all the bases (i.e., sets of eight mutually orthogonal rays) in the system. Because there are 2025 bases, and this is too large a number to display explicitly, we list just a few bases and give a simple set of rules (based on the symmetries of E8) for generating all the bases from the ones shown. It should be pointed out that the entire basis table can in fact be generated from any one of its elements by applying products of all possible powers of three basic symmetry operations of E8. The basis table is of central importance in this paper because all the parity proofs of E8 can be exhibited as subsets of it.\\

We say a few words about the broader area in which this work is set, to provide some perspective. After their initial discovery in two-qubit systems\cite{Peres1991,Kernaghan1994,Cabello1996,Pavicic2010,Waegell2011c}, parity proofs were discovered in three-\cite{KP1995,Waegell2012a} and higher-qubit\cite{DiVin,Waegell2013} systems, in systems of rays derived from the four-dimensional regular polytopes \cite{Waegell2011a,Waegell2011b,Waegell2014} and the root systems of exceptional Lie algebras\cite{ruugeE6} and, very recently, in a remarkably compact six-dimensional system of complex rays\cite{Lisonek2013} in connection with which an experiment has also been reported\cite{canas2014a}. Aside from revealing the many guises in which quantum contextuality can arise in spaces of different dimensionality, parity proofs are interesting because they have a variety of applications: they can be used to derive state-independent inequalities for ruling out noncontextuality\cite{Cabello2008, badziag,kirchmair,bartosik,amselem ,moussa} and Bell inequalities for identifying fully nonlocal correlations\cite{Aolita}; they have applications to quantum games\cite{Ambrosio}, quantum zero-error communication\cite{Cubitt}, quantum error correction\cite{Error,raussendorf} and the design of relational databases\cite{Abramsky2012}; they can be used to witness the dimension of quantum systems\cite{guhne2014}; and they underlie surprising phenomena such as the quantum pigeonhole effect\cite{aharonov2014,yu2014,rae2014}. Although the KS theorem is theoretically compelling, it has been argued\cite{mayer,kent,barrett} that the finite precision of real measurements nullifies practical attempts at verifying it. There is some debate about this matter\cite{mermin}, but it should be mentioned that methods of establishing contextuality that are not open to this objection have been proposed\cite{Klyachko,Liang,Bengtsson,YuOh}. Spekkens\cite{spekkens} has recently expanded upon the conditions that must be satisfied by realistic experiments that claim to rule out noncontextual ontological models. It has been argued in \cite{emerson} that contextuality is the source of the speedup in many quantum information protocols. This brief survey is far from complete, but it serves to show that the KS theorem and quantum contextuality are at the heart of many current research efforts.\\

The plan of this paper is as follows. Section \ref{sec:2} introduces the triacontagonal representation of the 120 rays derived from the root vectors of E8, and shows how their symmetries can be exploited to give an efficient construction of their basis table. Section \ref{sec:3} points out some interesting substructures within E8 that have been shown in the past to give rise to proofs of the KS theorem. Section \ref{sec:4} reviews the notion of a parity proof and shows how the triacontagonal representation of E8 facilitates the identification of subsets of its bases, of distinct symmetry types, that each house a large number of parity proofs. The significance of these smaller subsets is that they are far more easily searched for parity proofs than the entire system. After giving an overview of the parity proofs we found among the different subsets of bases, we present a few examples of the parity proofs that illustrate their important features. Section \ref{sec:5} concludes with a discussion of our results.

\section{\label{sec:2} The E8 system: rays and bases}

The 240 root vectors of E8 come in 120 pairs, with the members of each pair being the negatives of each other. Choosing just one member from each pair yields the 120 rays associated with E8. Each ray has eight real coordinates, which may be chosen in a variety of ways. We use a coordinatization due to Richter\cite{richter}, which differs from the one introduced earlier by Coxeter and Shepard\cite{leonardo}. Let $\omega = \exp(\frac{i\pi}{30})$ and let $\tau = \frac{1+\surd 5}{2}$ be the golden ratio. Define $a,b,c$ and $d$ as the positive numbers satisfying the equations \\

\hspace{5mm} $2a^{2}=1+3^{-1/2}5^{-1/4}\tau^{3/2} , \hspace{5mm} 2b^{2}=1+3^{-1/2}5^{-1/4}\tau^{-3/2}$ \\
\indent
\hspace{5mm} $2c^{2}=1-3^{-1/2}5^{-1/4}\tau^{-3/2} , \hspace{3mm} 2d^{2}=1-3^{-1/2}5^{-1/4}\tau^{3/2}$ .\\

\noindent
For any integer $n$, let $c_{n}=\omega^{n}+\omega^{-n}=2\cos(\frac{n\pi}{30})$ and define the quantities \\

\hspace{2mm} $r_{1}=a/c_{9}$ , \hspace{2mm} $r_{2}=b/c_{9}$ ,  \hspace{2mm} $r_{3}=c/c_{9}$ ,  \hspace{2mm} $r_{4}=d/c_{9}$ , \\
\indent
\hspace{2mm} $r_{5}=a/c_{3}$ , \hspace{2mm} $r_{6}=b/c_{3}$ ,  \hspace{2mm} $r_{7}=c/c_{3}$ ,  \hspace{2mm} $r_{8}=d/c_{3}$ . \\

\noindent
The 120 rays $|i\rangle$ ($i=1,\cdots, 120$) are then defined as\\

\noindent
$|n+1\rangle = (r_{1}\omega^{2n},r_{4}\omega^{22n},r_{6}\omega^{14n+1},r_{7}\omega^{26n+1})$ \hspace{23 mm} for $0\leq n \leq 14$\\
$|n+16\rangle = (r_{4}\omega^{2n},-r_{1}\omega^{22n},r_{7}\omega^{14n+1},-r_{6}\omega^{26n+1})$ \hspace{16 mm} for $0\leq n \leq 14$\\
$|n+23\rangle = (r_{7}\omega^{29+2n},-r_{6}\omega^{19+22n},-r_{1}\omega^{24+14n},r_{4}\omega^{18+26n})$ \hspace{3 mm} for $8\leq n \leq 14$\\
$|n+38\rangle = (r_{7}\omega^{29+2n},-r_{6}\omega^{19+22n},-r_{1}\omega^{24+14n},r_{4}\omega^{18+26n})$ \hspace{3 mm} for $0\leq n \leq 7$\\
$|n+38\rangle = (r_{6}\omega^{29+2n},r_{7}\omega^{19+22n},r_{4}\omega^{24+14n},r_{1}\omega^{18+26n})$ \hspace{9 mm} for $8\leq n \leq 14$\\
$|n+53\rangle = (r_{6}\omega^{29+2n},r_{7}\omega^{19+22n},r_{4}\omega^{24+14n},r_{1}\omega^{18+26n})$ \hspace{9 mm} for $0\leq n \leq 7$\\
$|n+61\rangle = (r_{8}\omega^{2n},-r_{5}\omega^{22n},-r_{3}\omega^{14n+1},r_{2}\omega^{26n+1})$ \hspace{16 mm} for $0\leq n \leq 14$\\
$|n+76\rangle = (r_{5}\omega^{2n},r_{8}\omega^{22n},-r_{2}\omega^{14n+1},-r_{3}\omega^{26n+1})$ \hspace{16 mm} for $0\leq n \leq 14$\\
$|n+83\rangle = (r_{2}\omega^{29+2n},r_{3}\omega^{19+22n},-r_{8}\omega^{24+14n},-r_{5}\omega^{18+26n})$ \hspace{3 mm} for $8\leq n \leq 14$\\
$|n+98\rangle = (r_{2}\omega^{29+2n},r_{3}\omega^{19+22n},-r_{8}\omega^{24+14n},-r_{5}\omega^{18+26n})$ \hspace{3 mm} for $0\leq n \leq 7$\\
$|n+98\rangle = (r_{3}\omega^{29+2n},-r_{2}\omega^{19+22n},r_{5}\omega^{24+14n},-r_{8}\omega^{18+26n})$ \hspace{3 mm} for $8\leq n \leq 14$\\
$|n+113\rangle = (r_{3}\omega^{29+2n},-r_{2}\omega^{19+22n},r_{5}\omega^{24+14n},-r_{8}\omega^{18+26n})$ \hspace{2 mm} for $0\leq n \leq 7$ \hspace{1 mm} ,\\

\noindent
with each ray being an 8-component column vector whose (real) components are given by the real and imaginary parts of the four complex numbers listed for it\footnote{For example, the components of the column vector $|3\rangle$, in the proper order, are $r_{1}\cos(\frac{2\pi}{15}),r_{1}\sin(\frac{2\pi}{15}),r_{4}\cos(\frac{22\pi}{15}),r_{4}\sin(\frac{22\pi}{15}),
r_{6}\cos(\frac{29\pi}{30}),r_{6}\sin(\frac{29\pi}{30}),r_{7}\cos(\frac{53\pi}{30}),r_{7}\sin(\frac{53\pi}{30}).$}. We will use $\langle i |$ to denote the 8-component row vector that is the transpose of $|i\rangle$.\\

Let us denote by the letters A,...,H each consecutive set of 15 rays (thus A denotes rays 1-15, B rays 16-30, etc.). If we add to each group of 15 rays all their negatives, we get groups of 30 vectors whose first two coordinates define the vertices of regular triacontagons in the plane, with the triacontagons corresponding to the eight letter groups being concentric to one another. This is just the triacontagonal representation of the roots of E8 (or of the vertices of Gosset's polytope $4_{21}$) mentioned in the introduction. Although the coordinates we have introduced for the rays are identical to those of Richter\cite{richter}, our numbering of the rays is a bit different from his (in essence, we have swapped some of his triacontagons and rotated some of them relative to the others for convenience in the presentation of some of our results).\\

A straightforward calculation shows that each of the 120 rays is orthogonal to 63 others and that the rays form 2025 bases. Each ray occurs in 135 bases and its only companions in these bases are the 63 other rays it is orthogonal to. We will denote this system of rays and bases by the symbol $120_{135}$-$2025_{8}$, with the subscript on the left indicating the multiplicity of each of the rays (i.e., the number of bases it occurs in) and that on the right the number of rays in each basis. The product of the numbers in the left half of the symbol equals the product on the right, as it should. The basis table of E8 (i.e., the complete list of all its bases) is \textit{saturated}, by which we mean that all the orthogonalities between its rays are represented in it. Because of this, the basis table is completely equivalent to the Kochen-Specker diagram of its rays\footnote{This is a graph whose vertices correspond to the rays and whose edges join vertices corresponding to orthogonal rays}. However it has the great advantage over the Kochen-Specker diagram that it is easy to interpret and work with.\\

Later we will encounter other systems of rays and bases having a high degree of symmetry, and the notation we have introduced above is easily modified to deal with such cases. For example, a system of 45 rays that forms 15 bases, with 30 of the rays being of multiplicity 2 and the other 15 of multiplicity 4 can be represented by the symbol $30_{2}15_{4}$-$15_{8}$ (again the sum of the products of each number on the left with its subscript equals the product of the number and its subscript on the right). The parity proofs we will present later, which are subsets of the basis table, can also be described by symbols of this kind.\\

We now present the basis table of E8. Figure 1 shows 15 bases that contain all 120 rays once each. The entire basis table can be derived from these 15 bases by permuting the rays in them in the manner we now describe. Let $V$ be the permutation of order 9 with the cycle decomposition\footnote{By the cycle (1 5 9 13 ... 11), we mean the permutation in which 1 goes to 5, 5 to 9, 9 to 13 ... and 11 to 1.} $V$ = (1 5 9 13 53 40 82 105 11)(2 91 90 55 28 42 54 119 49)(3 47 38 66 31 30 41 103 12)(4 10 51 89 117 106 87 27 36)(6 93 97 101 86 71 48 69 113)(7 14 79 67 33 29 64 32 100)(8 95 99 45 44 92 112 63 78)(15 104 34 46 109 77 118 107 85)(16 120 98 60 61 75 18 35 68)(17 73 20 24 59 96 58 57 94)(19 23 76 52 84 56 21 25 37)(22 74 116 108 115 72 62 50 70)(65 80 88 102 83 110 114 81 111)(26 43 39) and $W$ the permutation of order 15 with the cycle decomposition $W$ = (1 $\cdots$ 15)(16 $\cdots$ 30)(31 $\cdots$ 45)(46 $\cdots$ 60)(61 $\cdots$ 75)(76 $\cdots$ 90)(91 $\cdots$ 105)(106 $\cdots$ 120), where the dots signify all the integers between the two extremes. Let each basis in Figure 1 be assigned the label $(0,0,l)$, where the first two labels are fixed and the third varies, in integer steps, from $0$ to $14$. Then any other member of the basis table, which is assigned the label $(n,m,l)$ with $0\leq l,n \leq 14$ and $0 \leq m \leq 8$, can be generated by applying suitable powers of $W$ and $V$ to one of the bases in Fig.1, as described by the equation $(n,m,l) = W^{n}V^{m}(0,0,l)$. The number of bases that can be generated in this way is $15\cdot 9 \cdot 15 = 2025$, which is the entire basis table.\\

\begin{figure}[ht]
\centering 
\begin{tabular}{|c | c |} 
\hline 
Basis label & Basis  \\
\hline
(0,0,0) &  1 7  62  66  70  73 107 111 \\
(0,0,1) & 29 115 33	11	74	61	5	52 \\
(0,0,2) & 64 34	85	102	19	88	101	94 \\
(0,0,3) & 58 78	92	42	47	110	112	116\\
(0,0,4) & 55 117	80	51	96	106	41	108\\
(0,0,5) & 27 81	84	82	21	59	114	14\\
(0,0,6) & 10 23	38	103	37	56	53	113\\
(0,0,7) & 2	72	44	104	95	32	48	25\\
(0,0,8) & 17 89	9	76	54	26	119	77\\
(0,0,9) & 87 109	4	16	22	86	79	49\\
(0,0,10) & 100 31	120	45	60	57	13	30\\
(0,0,11) & 91	90	43	15	75	20	105	63\\
(0,0,12) & 99	71	36	28	93	39	6	50\\
(0,0,13) & 24	46	67	35	118	12	3	69\\
(0,0,14) & 83	8	40	97	98	18	65	68\\
\hline
\end{tabular}
\caption{Fifteen bases of the E8 system, involving the rays 1-120 once each. The first column shows the three-index label of each basis.}
\label{tab1} 
\end{figure}

In Figure \ref{tab2} we show the 9 blocks of bases obtained by applying all powers of $V$ to the block of Figure \ref{tab1}. The remaining blocks of bases can be obtained by applying powers of $W$ to these nine blocks. Since an application of $W$ amounts, for the most part, to increasing the ray numbers by one, these other blocks are easily written down.\\

\begin{figure}[h!]
\centering 
\begin{tabular}{|c | c |c |} 
\hline 
1	7	62	66	70	73	107	111 &	5	14	50	31	22	20	85	65 &	9	79	70	30	74	24	15	80\\
29	115	33	11	74	61	5	52	& 64	72	29	1	116	75	9	84 &	32	62	64	5	108	18	13	56\\
64	34	85	102	19	88	101	94	& 32	46	15	83	23	102	86	17 &	100	109	104	110	76	83	71	73\\
58	78	92	42	47	110	112	116	& 57	8	112	54	38	114	63	108 &	94	95	63	119	66	81	78	115\\
55	117	80	51	96	106	41	108	& 28	106	88	89	58	87	103	115	&   42	87	102	117	57	27	12	72\\
27	81	84	82	21	59	114	14	& 36	111	56	105	25	96	81	79	&   4	65	21	11	37	58	111	67\\
10	23	38	103	37	56	53	113	& 51	76	66	12	19	21	40	6	&   89	52	31	3	23	25	82	93\\
2	72	44	104	95	32	48	25	& 91	62	92	34	99	100	69	37	&   90	50	112	46	45	7	113	19\\
17	89	9	76	54	26	119	77	& 73	117	13	52	119	43	49	118	&   20	106	53	84	49	39	2	107\\
87	109	4	16	22	86	79	49	& 27	77	10	120	74	71	67	2	&   36	118	51	98	116	48	33	91\\
100	31	120	45	60	57	13	30	& 7	    30	98	44	61	94	53	41	&   14  41	60	92	75	17	40	103\\
91	90	43	15	75	20	105	63	& 90	55	39	104	18	24	11	78	&   55	28	26	34	35	59	1	8\\
99	71	36	28	93	39	6	50	& 45	48	4	42	97	26	93	70	&   44	69	10	54	101	43	97	22\\
24	46	67	35	118	12	3	69	& 59	109	33	68	107	3	47	113	&   96	77	29	16	85	47	38	6\\
83	8	40	97	98	18	65	68	& 110	95	82	101	60	35	80	16	&  114	99	105	86	61	68	88	120\\
\hline
\end{tabular}

\centering
\begin{tabular}{c c }
  & \\
\end{tabular}

\centering 
\begin{tabular}{|c | c |c |} 
\hline 
13	67	22	41	116	59	104	88&	53	33	74	103	108	96	34	102&40	29	116	12	115	58	46	83\\
100	50	32	9	115	35	53	21&	7	70	100	13	72	68	40	25&	14	22	7	53	62	16	82	37\\
7	77	34	114	52	110	48	20&	14	118	46	81	84	114	69	24&	79	107	109	111	56	81	113	59\\
17	99	78	49	31	111	8	72&	73	45	8	2	30	65	95	62&	20	44	95	91	41	80	99	50\\
54	27	83	106	94	36	3	62&	119	36	110	87	17	4	47	50&	49	4	114	27	73	10	38	70\\
10	80	25	1	19	57	65	33&	51	88	37	5	23	94	80	29&	89	102	19	9	76	17	88	64\\
117	84	30	47	76	37	105	97&	106	56	41	38	52	19	11	101&87	21	103	66	84	23	1	86\\
55	70	63	109	44	14	6	23&	28	22	78	77	92	79	93	76&	42	74	8	118	112	67	97	52\\
24	87	40	56	2	26	91	85&	59	27	82	21	91	43	90	15&	96	36	105	25	90	39	55	104\\
4	107	89	60	108	69	29	90&	10	85	117	61	115	113	64	55&	51	15	106	75	72	6	32	28\\
79	103	61	112	18	73	82	12&	67	12	75	63	35	20	105	3&	33	3	18	78	68	24	11	47\\
28	42	43	46	68	96	5	95&	42	54	39	109	16	58	9	99&	54	119	26	77	120	57	13	45\\
92	113	51	119	86	39	101	74&	112	6	89	49	71	26	86	116&63	93	117	2	48	43	71	108\\
58	118	64	120	15	38	66	93&	57	107	32	98	104	66	31	97&	94	85	100	60	34	31	30	101\\
81	45	11	71	75	16	102	98&	111	44	1	48	18	120	83	60&	65	92	5	69	35	98	110	61\\

\hline
\end{tabular}

\centering
\begin{tabular}{c c }
  & \\
\end{tabular}

\centering 
\begin{tabular}{|c | c |c |} 
\hline 
82	64	108	3	72	57	109	110&105	32	115	47	62	94	77	114&11	100	72	38	50	17	118	81\\
79	74	14	40	50	120	105	19&	67	116	79	82	70	98	11	23&	33	108	67	105	22	60	1	76\\
67	85	77	65	21	111	6	96&	33	15	118	80	25	65	93	58&	29	104	107	88	37	80	97	57\\
24	92	99	90	103	88	45	70&	59	112	45	55	12	102	44	22&	96	63	44	28	3	83	92	74\\
2	10	81	36	20	51	66	22&	91	51	111	4	24	89	31	74&	90	89	65	10	59	117	30	116\\
117	83	23	13	52	73	102	32&	106	110	76	53	84	20	83	100&87	114	52	40	56	24	110	7\\
27	25	12	31	56	76	5	71&	36	37	3	30	21	52	9	48&	4	19	47	41	25	84	13	69\\
54	116	95	107	63	33	101	84&	119	108	99	85	78	29	86	56&	49	115	45	15	8	64	71	21\\
58	4	11	37	55	26	28	34&	57	10	1	19	28	43	42	46&	94	51	5	23	42	39	54	109\\
89	104	87	18	62	93	100	42&	117	34	27	35	50	97	7	54&	106	46	36	68	70	101	14	119\\
29	47	35	8	16	59	1	38&	64	38	68	95	120	96	5	66&	32	66	16	99	98	58	9	31\\
119	49	43	118	98	94	53	44&	49	2	39	107	60	17	40	92&	2	91	26	85	61	73	82	112\\
78	97	106	91	69	39	48	115&8	101	87	90	113	26	69	72&	95	86	27	55	6	43	113	62\\
17	15	7	61	46	30	41	86&	73	104	14	75	109	41	103	71&	20	34	79	18	77	103	12	48\\
80	112	9	113	68	60	114	75&	88	63	13	6	16	61	81	18&	102	78	53	93	120	75	111	35\\
\hline
\end{tabular}
\caption{The nine blocks of bases obtained by applying all powers of $V$, from 0 to 8, to the block of Figure \ref{tab1}. The powers increase as one goes from left to right and up to down. Applying $V$ to the block at the bottom right gives back the block at the top left, which is just the block of Fig.\ref{tab1}.}
\label{tab2}
\end{figure}

The construction we have given of the basis table can be compressed even further by introducing a permutation of order 15, which we will term $U$, whose cycle decomposition is given by the eight columns of numbers obtained by aligning the bases in Fig.\ref{tab1} and reading down them vertically\footnote{To be explicit, $U$ =(1 29 64 $\cdots$ 83)(7 115 34 $\cdots$ 8) $\cdots$ (111 52 94 $\cdots$ 68), where there are 8 cycles and each consists of 15 numbers.}. Then all the bases can be generated from the first basis of Fig.\ref{tab1} by applying powers of the permutations $U,V$ and $W$ to it, as described by the equation

\begin{equation}
(n,m,l) = W^{n}V^{m}U^{l}(0,0,0) \hspace{2mm} ,
\label{eq1}
\end{equation} \

\noindent
with $0 \leq l \leq 14, 0 \leq m \leq 8$ and $0 \leq n \leq 14$. This procedure works even if an arbitrary basis is substituted for $(0,0,0)$ as the seed basis. The three-index label $(n,m,l)$ serves as a convenient shorthand for the basis if one wishes to avoid listing all its rays. \\

Instead of describing $U,V$ and $W$ by the permutations they perform on the rays, one can represent them by the $8 \times 8$ orthogonal matrices
\begin{equation}
U = -|29\rangle \langle 1|-|115\rangle \langle 7|-|33\rangle \langle 62|+|11\rangle \langle 66|+|74\rangle \langle 70|+|61\rangle \langle 73|-|5\rangle \langle 107|-|52\rangle \langle 111|
\label{eq2a}
\end{equation}
\begin{equation}
V = |5\rangle \langle 1|+|14\rangle \langle 7|-|50\rangle \langle 62|-|31\rangle \langle 66|-|22\rangle \langle 70|-|20\rangle \langle 73|-|85\rangle \langle 107|+|65\rangle \langle 111|
\label{eq2b}
\end{equation}
and
\begin{equation}
W = |2\rangle \langle 1|+|8\rangle \langle 7|+|63\rangle \langle 62|+|67\rangle \langle 66|+|71\rangle \langle 70|+|74\rangle \langle 73|+|108\rangle \langle 107|+|112\rangle \langle 111| \hspace{2mm} ,
\label{eq2c}
\end{equation}
as one can easily check by applying them to the column vectors representing the rays and verifying that they produce the desired permutations.\\

We can construct all the symmetries of E8 by looking at the mappings of a fixed basis on to all the bases of the system. Let $x_{1} x_{2} \cdots x_{8}$ and $y_{1} y_{2} \cdots y_{8}$ be two bases and let $y'_{1} y'_{2} \cdots y'_{8}$ be some permutation of the numbers in the latter. Consider the orthogonal transformation
\begin{equation}
T= (-1)^{n_{1}}|y'_{1}\rangle \langle x_{1}|+(-1)^{n_{2}}|y'_{2} \rangle \langle x_{2}|+\cdots+(-1)^{n_{8}}|y'_{8}\rangle \langle x_{8}| \hspace{2mm} ,
\label{eq3}
\end{equation}
where $n_{i}\in (0,1)$ for $1 \leq i \leq 8$. Keeping the basis $x_{1} x_{2} \cdots x_{8}$ fixed and letting $y_{1} y_{2} \cdots y_{8}$ vary, the number of transformations of the form (\ref{eq3}) that one can construct is the number of possibilities for the variable basis (2025) times the number of permutations of the variable basis labels ($8!$) times the number of possibilities for the signs of the terms ($2^{8}$). However an investigation shows that only $1/30$ of the $8!$ permutations of the basis labels lead to symmetries of E8, so the total number of its symmetries is $2025\cdot\frac{8!}{30}\cdot2^{8}=696729600$, which equals the number of $192\cdot 10!$ given by Coxeter\cite{coxeter} as the order of the symmetry group of Gosset's polytope $4_{21}$.

\section{\label{sec:3}Substructures within the E8 system}

The E8 system contains a number of interesting substructures that yield proofs of the KS theorem. These substructures have all been studied in the past, and we discuss each of them briefly. \\

Two interesting substructures are the Lie algebras E7 and E6, whose roots/rays can be exhibited as subsets of the roots/rays of E8. The rays of E7 are simply the 63 rays orthogonal to any ray of E8; these rays lie in a 7-dimensional space, and if one adjoins to them all their negatives, one gets the 126 roots of E7. The 63 rays of E7 form 135 bases, with each ray occurring in 15 bases. This system can be described by the symbol $63_{15}$-$135_{7}$, and it is saturated. The basis table of E7 is easily extracted from that of E8 by picking out the 135 bases involving a particular ray and then dropping that ray from these bases. This construction shows that the symmetry group of E7 is a subgroup of index 240 of that of E8; thus its order is $192 \cdot 10!/240$, which agrees with the figure of $8\cdot 9!$ given by Coxeter for the order of the symmetry group of the associated 7-dimensional polytope $3_{21}$. The rays of E7 can be used to give a proof of the KS theorem. Because the system is saturated, the proof requires showing that it is impossible to assign noncontextual $0/1$ values to the rays in such a way that each of the 135 bases has just a single ray assigned the value 1 in it, and this is easily done using a ``proof-tree'' argument\footnote {The argument is a \textit{reductio ad absurdum} one: one assumes that a noncontextual value assignment exists and then shows that it leads to a contradiction. Since E7 has a symmetry group that is transitive on its rays, one can begin, without loss of generality, by assigning an arbitrary ray the value 1. This forces all rays orthogonal to that ray to have the value 0, and one then finds that a basis with three rays having the value 0 appears. Assigning one of the remaining rays in this basis the value 1 forces a basis with five rays having the value 0 to appear. However assigning either of the remaining rays in this basis the value 1 forces a basis with all its rays assigned the value 0 to appear, which is not allowed. To avoid this conflict, one must proceed backwards along the chain and make alternative choices for the rays assigned the value 1 at every earlier step of the argument and see if any of these alternative possibilities leads to a valid value assignment. One then finds that all the alternatives lead to a situation in which at least one basis has all its rays assigned the value 0, showing that a valid value assignment does not exist and proving the KS theorem. The ``proof-tree'' leading to this contradiction has eight branches, with each branch leading to a contradiction at the fourth step.}. An alternative proof of the KS theorem based on the rays of E7 has been given by Ruuge\cite{ruugeE6}.\\

If one picks the 36 rays orthogonal to any two nonorthogonal rays of E8, one gets the rays of the E6 system, and the roots of E6 are these 36 rays along with  their negatives. The rays of E6 do not form even a single basis (i.e., a set of six mutually orthogonal rays) and so do not yield a proof of the KS theorem. This was pointed out by Ruuge\cite{ruugeE6}, who discussed how two rotated copies of E6 could be superposed to give a system of rays that yields a proof of the KS theorem.\\

Another interesting subsystem of E8 is what we will term a Kernaghan-Peres (KP) set\cite{KP1995}. Such a set consists of 40 rays that form 25 bases, with each ray occurring in five bases, so that its symbol is $40_{5}$-$25_{8}$. Kernaghan and Peres\cite{KP1995} constructed such a set as the simultaneous eigenstates of five sets of mutually commuting observables of a system of three qubits. The caption to Fig.\ref{tab3} explains how KP sets can be extracted from the bases of E8 and the figure gives an example of a set constructed using this procedure. The caption to Fig.\ref{tab3a} gives a simple procedure for obtaining all the parity proofs in a KP set\cite{Waegell2012a} and the figure gives one example of each of the three types of parity proofs contained in the KP set of Fig.\ref{tab3} (see the beginning of Sec. \ref{sec:4} for an explanation of the notion of a parity proof). Finally, Fig.\ref{tab3b} gives an example of a ``pseudo'' KP set that closely resembles a KP set but is not one, and in fact yields no proofs of the KS theorem.

\begin{figure}[ht]
\centering 
\begin{tabular}{|c | c |c |} 
\hline 
1	7	62	66	70	73	107	111&	1	64	94	7	66	62	34	19	&	101	102	70	111	73	88	85	107\\
29	115	33	11	74	61	5	52&	    52	1	107	61	73	7	33	29	&	5	11	66	70	74	62	111	115\\
64	34	85	102	19	88	101	94&		62	92	116	1	47	42	70	107	&	78	110	7	111	66	58	112	73\\
58	78	92	42	47	110	112	116&	66	38	1	10	23	73	103	70	&	56	7	111	53	37	113	62	107\\
10	23	38	103	37	56	53	113&	29	115	47	78	58	33	11	116	&	112	52	42	74	61	92	5	110\\
								&	5	94	29	88	64	61	85	115	&	19	33	34	52	74	102	11	101\\
								&	52	5	11	29	23	38	37	56	&	53	113	74	115	10	33	61	103\\
								&	88	92	116	101	64	34	58	110	&	19	94	112	78	42	85	102	47\\
								&	94	34	103	23	88	37	53	102	&	10	64	19	38	113	56	101	85\\
								&	113	58	37	112	92	103	47	38	&	110	42	116	78	23	10	53	56\\
\hline
\end{tabular}
\caption{A Kernaghan-Peres (KP) set can be extracted from the bases of E8 by choosing any five (``seed'') bases from one of the blocks in Fig.\ref{tab2} (or a block obtained from one of these blocks by applying a power of $W$ to it) and supplementing them by 20 bases, chosen from the full set of 2025, that each have four rays from one seed basis and four rays from another. The 20 added bases always come in 10 complementary pairs, with the members of each pair originating in the same pair of seed bases and having no rays in common. In order that this construction gives rise to a KP set, it is necessary that each ray occurs once with 17 other rays and thrice with 6 other rays in the five bases in which it occurs. Shown above is a KP set constructed according to this procedure, with its five seed bases shown in the first column and its 10 pairs of complementary bases in the second and third columns; the seed bases were picked from the block in the top left corner of Fig.\ref{tab2} and each pair of complementary bases is shown on a line. It can be checked that this set of rays and bases has the symbol $40_{5}$-$25_{8}$ and that each of its rays has the pattern of companions stated earlier.}
\label{tab3}
\end{figure}

\begin{figure}[ht]
\centering 
\begin{tabular}{|c | c |c |} 
\hline 
1	64	94	7	66	62	34	19   & 1 64	94	7	66	62	34	19  &  1	64	94	7	66	62	34	19 \\
52	1	107	61	73	7	33	29   &52 1	107	61	73	7	33	29  &52	1	107	61	73	7	33	29 \\
62	92	116	1	47	42	70	107  &62 92	116	1	47	42	70	107 & 62	92	116	1	47	42	70	107 \\
66	38	1	10	23	73	103	70   &66 38	1	10	23	73	103	70 &56	7	111	53	37	113	62	107 \\
29	115	47	78	58	33	11	116  &29 115 47	78	58	33	11	116 &29	115	47	78	58	33	11	116 \\
5	94	29	88	64	61	85	115  &5	94	29	88	64	61	85	115 &5	94	29	88	64	61	85	115 \\
52	5	11	29	23	38	37	56   &52 5	11	29	23	38	37	56& 53	113	74	115	10	33	61	103 \\
88	92	116	101	64	34	58	110  &88 92	116	101	64	34	58	110&88	92	116	101	64	34	58	110 \\
94	34	103	23	88	37	53	102  &94 34	103	23	88	37	53	102& 94	34	103	23	88	37	53	102 \\
110	42	116	78	23	10	53	56   &113 58	37	112	92	103	47	38& 113	58	37	112	92	103	47	38\\
64	34	85	102	19	88	101	94   &64 34	85	102	19	88	101	94& 1	7	62	66	70	73	107	111 \\
                                 &58 78	92	42	47	110	112	116&29	115	33	11	74	61	5	52 \\
                                 &10 23	38	103	37	56	53	113&64	34	85	102	19	88	101	94 \\
                                 &                             &58	78	92	42	47	110	112	116\\
                                 &                             &10	23	38	103	37	56	53	113\\
\hline
\end{tabular}
\caption{Parity proofs can be extracted from any KP set by picking one member from each of the 10 pairs of complementary bases and supplementing them with the needed seed bases. There are three types of parity proofs that can be constructed in this way, and they involve the addition of one, three or five seed bases. These proofs have the symbols $28_{2}8_{4}$-$11_{8}$, $24_{2}14_{4}$-$13_{8}$ and $20_{2}20_{4}$-$15_{8}$, and one example of each, constructed from the KP set of Fig.\ref{tab3}, is shown in the three columns above (with the seed bases always shown at the end). The 10 bases in the first step of the construction can be picked in $2^{10}=1024$ ways, and all of them can be extended into valid parity proofs.}
\label{tab3a}
\end{figure}

\begin{figure}[ht]
\centering 
\begin{tabular}{|c | c |c |} 
\hline 
1	7	62	66	70	73	107	111	&	1	64	94	7	66	62	34	19	&	101	102	70	111	73	88	85	107 \\
64	34	85	102	19	88	101	94	&	26	76	9	119	73	70	66	1	&	7	111	54	89	77	107	62	17 \\
17	89	9	76	54	26	119	77	&	28	50	71	39	107	1	7	73	&	66	93	6	70	36	99	111	62 \\
99	71	36	28	93	39	6	50	&	18	83	1	111	65	62	73	8	&	66	40	70	97	107	98	7	68 \\
83	8	40	97	98	18	65	68	&	89	102	19	9	76	17	88	64	&	26	101	119	85	34	77	94	54 \\
								&	85	50	34	99	28	19	6	88	&	93	64	94	36	101	102	71	39 \\
								&	101	68	34	18	8	88	97	64	&	19	102	94	85	98	83	40	65 \\
								&	119	54	39	36	50	17	9	99	&	77	26	93	71	89	28	6	76 \\
								&	54	76	40	8	26	97	83	17	&	9	98	18	77	68	65	89	119 \\
								&	99	83	36	8	68	28	98	71	&	40	65	6	50	97	39	93	18 \\

\hline
\end{tabular}
\caption{The construction that gave rise to the KP set of Fig.\ref{tab3} also leads to many ``pseudo'' KP sets, like the one above. Although a $40_{5}$-$25_{8}$ set, it is not a KP set because each of its rays occurs four times with a particular companion in five bases, which never happens in a KP set. Further, this ``pseudo'' KP set does not yield a proof of the KS theorem because it is possible to make valid noncontextual 0/1 assignments to its rays in many ways; one such assignment consists of assigning 1's to the rays 1,85,17,93 and 68 and 0's to all the others.}
\label{tab3b}
\end{figure}

\section{\label{sec:4}Parity proofs in the E8 system}

We will say that a set of projectors in a Hilbert space of even dimension furnishes a parity proof of the KS theorem if the projectors form an odd number of bases in such a way that each projector occurs in an even number of the bases (a basis is any set of mutually orthogonal projectors that sums to the identity, and we will allow for the possibility that the projectors are not all of the same rank). Such a set of projectors proves the KS theorem because it is impossible to assign noncontextual $0/1$ values to them in such a way that the sum of the values assigned to the projectors in any basis is always 1. Because an even-odd conflict makes this assignment impossible, we refer to this type of proof as a parity proof. In \cite{lisonek1} it was pointed out that the E8 system has $2^{1940}$ parity proofs in it, but no examples of such proofs were given. In this section we would like to describe a straightforward method we used to discover a large number of these proofs, and then present a few examples of them. These proofs are far more numerous and varied than those in the KP sets we know to be contained in the E8 system.\\

We will discuss only \textit{critical} parity proofs, where by a critical proof we mean one that ceases to provide a proof of the KS theorem if even a single basis is dropped from it\footnote{Dropping a single basis from a parity proof leaves an even number of bases, which can never provide a parity proof of the KS theorem. However the reduced system may not admit a valid noncontextual value assignment to its rays, and so provide a proof of the KS theorem. If this happens, the original parity proof would not be deemed critical.}. We restrict ourselves to critical proofs to avoid redundancy, since many noncritical proofs can often be reduced to the same critical proof. We present just a few of the more striking critical proofs we found from among the staggeringly large number that exist.\\

In Sec. \ref{sec:2} we introduced the letters A to H for each consecutive set of 15 rays of E8. These letters can be used to attach an 8-letter label to each basis. For example, the basis 1 7 62 66 70 73 107 111 would have the label AAEEEEHH. Viewed in terms of their labels (which specify how the rays of a basis are distributed over the triacontagons of E8), the bases fall into 33 families with distinct triacontagon profiles. We made the important discovery that if one looks at only the bases of a particular family, the parity proofs housed by them could be unearthed with relatively little effort. In the following subsections we discuss the parity proofs we found in a few of the families.

{\bf \subsection{\label{subsec:type1} Type 1 Bases and their parity proofs }}

Let us term bases with the profile AAEEEEHH, BBFFFFGG, CCEEGGGG or DDFFHHHH as Type 1 bases. The 15 bases with each of these profiles give a parity proof that can be characterized by the symbol $15_{4}30_{2}$-$15_{8}$. Figure \ref{tab4} shows the proof given by the bases with profile AAEEEEEHH (the proofs given by the other three profiles are very similar).\\

The Type 1 bases give rise to two other types of parity proofs if bases of different profiles can be combined. These proofs are characterized by the symbols $15_{4}70_{2}$-$25_{8}$ and $45_{4}50_{2}$-$35_{8}$ and there are 12 versions of each (all structurally identical, but involving different rays). The properties of the three different types of parity proofs made up only of Type 1 bases are summarized in the first row of Figure \ref{tab5}. \\

\begin{figure}[ht]
\centering 
\begin{tabular}{|c | c |c |} 
\hline 
1	7	62	66	70	73	107	111&6	12	67	71	75	63	112	116&11	2	72	61	65	68	117	106\\
2	8	63	67	71	74	108	112&7	13	68	72	61	64	113	117&12	3	73	62	66	69	118	107\\
3	9	64	68	72	75	109	113&8	14	69	73	62	65	114	118&13	4	74	63	67	70	119	108\\
4	10	65	69	73	61	110	114&9	15	70	74	63	66	115	119&14	5	75	64	68	71	120	109\\
5	11	66	70	74	62	111	115&10	1	71	75	64	67	116	120&15	6	61	65	69	72	106	110\\
\hline
\end{tabular}
\caption{A $15_{4}30_{2}$-$15_{8}$ parity proof made up of 15 bases with profile AAEEEEHH. The rays 1-15 and 106-120 have multiplicity 2, while rays 61-75 have multiplicity 4.}
\label{tab4}
\end{figure}

\begin{figure}[ht]
\centering 
\begin{tabular}{|c | c |c |} 
\hline 
Basis Type & Parity Proofs  &   Number  \\
\hline
Type 1 & $15_{4}30_{2}$-$15_{8}$ & 4\\
(AAEEEEHH, & $15_{4}70_{2}$-$25_{8}$ & 12\\
BBFFFFGG, & $45_{4}50_{2}$-$35_{8}$ & 12\\
CCEEEEGG, & & \\
DDFFHHHH) & & \\
\hline
Type 2 & $36_{2}$-$9_{8}$ & 20\\
(AABBEEFF, & $60_{2}$-$15_{8}$ & 24\\
CCDDGGHH) & & \\
\hline
 & $60_{2}$-$15_{8}$, $52_{2}8_{2}$-$17_{8}$, $16_{4}44_{2}$-$19_{8}$, $1_{6}14_{4}45_{2}$-$19_{8}$,
$2_{6}12_{4}46_{2}$-$19_{8}$ & \\
Type 3 & $24_{4}36_{2}$-$21_{8}$, $1_{6}22_{4}37_{2}$-$21_{8}$, $2_{6}20_{4}38_{2}$-$21_{8}$, $3_{6}18_{4}39_{2}$-$21_{8}$, $4_{6}16_{4}40_{2}$-$21_{8}$ & \\
(EEFFGGHH)  & $6_{6}12_{4}42_{2}$-$21_{8}$, $32_{4}28_{2}$-$23_{8}$, $1_{6}30_{4}29_{2}$-$23_{8}$, $2_{6}28_{4}30_{2}$-$23_{8}$, $3_{6}26_{4}31_{2}$-$23_{8}$ & 700,326 \\
  & $4_{6}24_{4}32_{2}$-$23_{8}$, $5_{6}22_{4}33_{2}$-$23_{8}$, $6_{6}20_{4}34_{2}$-$23_{8}$, $7_{6}18_{4}35_{2}$-$23_{8}$, $8_{6}16_{4}36_{2}$-$23_{8}$ & \\
\hline
Type 4 &$2_{4}32_{2}$-$9_{8}$, $36_{2}$-$9_{8}$, $8_{4}28_{2}$-$11_{8}$, $7_{4}30_{2}$-$11_{8}$, $13_{4}26_{2}$-$13_{8}$ &   \\
(AABBCCDD) &$2_{6}15_{4}24_{2}$-$15_{8}$, $1_{8}2_{6}9_{4}32_{2}$-$15_{8}$, $1_{8}3_{6}11_{4}33_{2}$-$17_{8}$, $12_{6}24_{4}24_{2}$-$27_{8}$  & \\
\hline
\end{tabular}
\caption{Parity proofs in E8. For each of the classes of bases in the first column, the second column lists the symbols of the parity proofs that exist and the third column the number of versions of each of the proofs. The listings for Type 1 and Type 2 bases are complete. For Type 3 bases, the listing in the second column is complete but only the total count of all the proofs has been included in the third column. For Type 4 proofs, only nine of the over hundreds of different types of proofs we found are listed. The left halves of the proof symbols in the second column indicate the numbers of rays of different multiplicities present in the proofs, with no attempt being made to associate the rays with projectors of different ranks. Figs. \ref{tab6}, \ref{tab8} and \ref{tab9} give examples of proofs in which this distinction is made.}
\label{tab5}
\end{figure}

{\bf \subsection{\label{subsec:type2} Type 2 Bases and their parity proofs }}

We will term bases with the profile AABBEEFF or CCDDGGHH as Type 2 bases. There are 30 bases with each of these profiles, and therefore 60 Type 2 bases in all. These bases contain just the two types of parity proofs shown in the second row of Figure \ref{tab5}.\\

Figure \ref{tab6} shows two $36_{2}$-$9_{8}$ proofs of this class that seem very similar at first sight, but are subtly different from one another. While both proofs involve 36 rays that occur two times each over 9 bases, the proof on the left always has the following pairs of rays occur together over the bases: (1,66), (7,62), (16,86), (22,87), (9,84), (19,64), (21,76), (14,89), (17,82), (2,72), (11,61), (27,77), (29,74), (26,81), (4,79), (12,67), (24,69) and (6,71). Thus each of these pairs of rays can be regarded as defining a two-dimensional subspace, or a rank-2 projector, and the proof can be reinterpreted as involving 18 rank-2 projectors that each occur twice over nine bases; this situation can be captured in the symbol $18^{2}_{2}$-$9_{4}$, where the superscript on the left indicates the rank of the projectors and the subscript their multiplicity, and the subscript on the right that each of the 9 bases is made up of four rank-2 projectors. For the proof on the right, 18 of the rays can be paired into the rank-2 projectors (1,16), (62,87), (75,80), (70,90), (82,72), (65,85), (67,77), (11,26) and (6,21), while the remaining 18 rays are associated with rank-1 projectors; thus the symbol of this proof can be written as $9^{2}_{2}18^{1}_{2}$-$9_{6}$, with the superscripts and subscripts on the left having the same meaning as before and the subscript on the right indicating that each basis is made up of six projectors (four of rank-1 and two of rank-2).\\

\begin{figure}[ht]
\centering 
\begin{tabular}{|c | c |c |} 
\hline 
1	7	16	22	62	66	86	87& &1 7	16	22	62	66	86	87  \\
9	1	19	21	64	66	84	76& &10 1	25	16	71	75	80	81  \\
7	14	17	19	62	64	82	89& &15 7	25	27	70	72	90	82 \\
2	9	27	29	72	74	77	84& &10 2	20	22	65	67	85	77 \\
11	2	26	17	72	61	81	82& &11 2	26	17	72	61	81	82 \\
4	11	29	16	74	61	79	86& &5 11	20	26	66	70	90	76  \\
12	4	22	24	67	69	87	79& &5 12	30	17	75	62	80	87  \\
6	12	21	27	67	71	76	77& &6 12	21	27	67	71	76	77  \\
14	6	24	26	69	71	89	81& &15 6	30	21	61	65	85	86  \\
\hline
\end{tabular}
\caption{Two parity proofs made up exclusively of Type 2 bases with the profile AABBEEFF or CCDDGGHH. The proof on the left involves only rank-2 projectors and is characterized by the symbol $18^{2}_{2}$-$9_{4}$, while the proof on the right involves a mixture of rank-2 and rank-1 projectors and is characterized by the symbol $9^{2}_{2}18^{1}_{2}$-$9_{6}$ (see text for explanation). If one ignores the distinction between rank-1 and rank-2 projectors and focuses only on the rays, both proofs can be described by the common symbol $36_{2}$-$9_{8}$ (indicating that there are 36 rays that each occur twice over the nine bases).}
\label{tab6}
\end{figure}

{\bf \subsection{\label{subsec:type3} Type 3 Bases and their parity proofs }}

Type 3 bases are those with the profile EEFFGGHH. There are 45 such bases involving 60 rays, and they form a $60_{6}$-$45_{8}$ system. Despite their small number, the bases of this system are a fecund lot and give rise to 20 different types of parity proofs, each of which can come in hundreds to thousands of versions. The symbols of all the possible proofs are shown in the third row of Figure \ref{tab5}. When the different versions of each of the proofs are taken into account, the total number of distinct proofs is 700,326. Figure \ref{tab7} shows a $6_{6}12_{4}42_{2}$-$21_{8}$ proof of this class involving rays of multiplicity 6,4 and 2.\\

\begin{figure}[ht]
\centering 
\begin{tabular}{|c | c |c |} 
\hline 
61	63	81	88	105	94	114	115&68	70	88	80	97	101	106	107&67	70	82	85	98	99	119	108\\
62	64	82	89	91	95	115	116&69	71	89	81	98	102	107	108&65	72	90	77	103	94	118	109\\
63	65	83	90	92	96	116	117&72	74	77	84	101	105	110	111&66	73	76	78	104	95	119	110\\
64	66	84	76	93	97	117	118&73	75	78	85	102	91	111	112&69	61	79	81	92	98	107	113\\
65	67	85	77	94	98	118	119&62	65	77	80	93	94	114	118&71	63	81	83	94	100	109	115\\
66	68	86	78	95	99	119	120&63	66	78	81	94	95	115	119&74	66	84	86	97	103	112	118\\
67	69	87	79	96	100	120	106&66	69	81	84	97	98	118	107&75	67	85	87	98	104	113	119\\
\hline
\end{tabular}
\caption{A $6_{6}12_{4}42_{2}$-$21_{8}$ parity proof made up of 21 Type 3 bases with the profile EEFFGGHH. The rays of multiplicity 6 are 66,81,94,98,118,119, those of multiplicity 4 are 63,65,67,69,77,78,84,85,95,97,107,115, and all the other rays have multiplicity 2.}
\label{tab7}
\end{figure}

{\bf \subsection{\label{subsec:type3} Type 4 Bases and their parity proofs }}

Type 4 bases have the profile AABBCCDD. There are 75 such bases involving 60 rays, and they form a $60_{10}$-$75_{8}$ system. We have found over 400 different types of parity proofs in this system, with each coming in anywhere from scores to thousands of versions. We show just nine of these proofs in the last row of Figure \ref{tab5}. There are no critical proofs with more than 27 bases in this class. The number of proofs in this class greatly exceeds those in the previous three classes combined. Figure \ref{tab8} shows a $36_{2}$-$9_{8}$  proof of this class and Fig. \ref{tab8a} shows a rather unusual proof consisting entirely of rank-2 projectors.\\

\begin{figure}[ht]
\centering 
\begin{tabular}{|c |} 
\hline 
1	7	19	25	33	34	48	52\\\
7	13	25	16	39	40	54	58\\\
10	1	28	19	42	43	57	46\\
3	10	18	16	40	37	46	49\\
5	12	20	18	42	39	48	51\\
11	3	26	24	33	45	54	57\\
4	11	26	28	34	37	58	55\\
13	5	20	22	43	31	52	49\\
12	4	22	24	45	31	51	55\\
\hline
\end{tabular}
\caption{A $36_{2}$-$9_{8}$ parity proof made up of 9 Type 4 bases with the profile AABBCCDD. Twenty of the rays can be grouped into the rank-2 projectors (1,19), (7,25), (16,40), (10,46), (5,20), (12,51), (11,26), (22,31), (24,45) and (4,55), while the remaining rays define 16 rank-1 projectors. Also, eight of the bases involve 6 projectors (two of rank-2 and four of rank-1) while the remaining basis involves four rank-2 projectors. Thus a more descriptive symbol for this proof is $10^{2}_{2}16^{1}_{2}$-$8_{6}1_{4}$.}
\label{tab8}
\end{figure}

\begin{figure}[ht]
\centering 
\begin{tabular}{|c |} 
\hline 
1	7	19	25	33	34	48	52\\
10	1	28	19	42	43	57	46\\
13	4	16	22	45	31	60	49\\
8	14	26	20	41	42	53	57\\
11	2	29	23	44	45	56	60\\
14	5	17	26	32	33	59	48\\
15	7	30	28	37	34	58	46\\
8	15	30	17	38	41	47	59\\
4	11	29	16	37	38	58	47\\
10	2	20	22	43	44	49	53\\
13	5	23	25	31	32	52	56\\
\hline
\end{tabular}
\caption{A parity proof made up of 11 Type 4 bases with the profile AABBCCDD. It consists of 44 rays that each occur twice over the 11 bases, so its symbol is $44_{2}$-$11_{8}$. However a more careful examination shows that the rays can be paired into the 22 rank-2 projectors (1,19), (2,44), (4,16), (5,32), (7,34), (25,52), (33,48), (17,59), 14,26), (15,30), (11,29), (37,58), (38,47), (8,41), (28,46), (23,56), (13,31), (45.60), (22,49), (10,43), (20,53) and (42,57) that each occur twice over the bases. Thus a more descriptive symbol for this proof would be $22^{2}_{2}$-$11_{4}$, with the subscript of 4 on the right indicating that there are four rank-2 projectors in each basis.}
\label{tab8a}
\end{figure}

We end by presenting a proof of this class, in Fig.\ref{tab9}, that involves 34 rays (2 of multiplicity four and 32 of multiplicity two) that occur over 9 bases. This proof is more economical than the best proofs found earlier in 8 dimensions, which involve 36 rays occurring an even number of times over 11 bases\cite{KP1995} or 81 bases\cite{cabello2005} or 9 bases\cite{Waegell2012a} . As explained in the caption to Fig.\ref{tab9}, this proof can also be interpreted as involving 26 rank-1 projectors and 4 rank-2 projectors that occur over 9 bases.\\

\begin{figure}[ht]
\centering 
\begin{tabular}{|c |} 
\hline 
1	7	19	25	33	34	48	52\\
4	10	22	28	36	37	51	55\\
7	13	25	16	39	40	54	58\\
10	1	28	19	42	43	57	46\\
3	10	18	16	40	37	46	49\\
5	12	20	18	42	39	48	51\\
12	4	27	25	34	31	55	58\\
3	10	25	27	33	36	57	54\\
13	5	20	22	43	31	52	49\\
\hline
\end{tabular}
\caption{A $32_{2}2_{4}$-$9_{8}$ parity proof made up of 9 Type 4 bases with the profile AABBCCDD, with rays 25 and 10 being of multiplicity four and all the others of multiplicity two. However the pairs of rays (1,19), (4,55), (5,20) and (16,40) always occur together over the bases, with each pair occurring twice. Interpreting these pairs as rank-2 projectors and the remaining 26 rays as rank-1 projectors allows us to attach the more descriptive symbol $4^{2}_{2}2^{1}_{4}24^{1}_{2}$-$8_{7}1_{8}$ to this proof, where the subscripts in the second half of the symbol indicate that there are eight bases of 7 projectors (with 6 being of rank-1 and 1 of rank-2) and one basis of 8 projectors (with all being of rank-1).}
\label{tab9}
\end{figure}

\section{\label{sec:5}Discussion}

We pointed out at the end of Sec. \ref{sec:3} that the bases of E8 have 33 different triacontagon profiles. Our survey of parity proofs in Sec. \ref{sec:4} covered just four of these profiles, so it is clear that we have left the vast majority of the proofs untouched. The basis table of E8 presented in this paper serves a convenient template for displaying all the proofs in this gargantuan system.\\

It is interesting that the triacontagonal representations of both the 600-cell and Gosset's polytope lead to some of the simplest parity proofs contained in them. In the case of the 600-cell, the vertices project into four triacontagons, with two of the triacontagons uniting to yield a parity proof of 15 bases and the other two triacontagons yielding the complementary proof (i.e, one involving all the rays not present in the earlier proof). In the case of Gosset's polytope, the vertices project into eight triacontagons, and one can construct a parity proof (actually four different proofs) by picking out 15 bases that each span all the triacontagons in the same way. The great virtue of the triacontagonal representation for Gosset's polytope (or E8) is, of course, that it allows the bases to be organized into smaller families that are more easily searched for their parity proofs. Although we have unearthed only a tiny fraction of the parity proofs present in E8, their variety and intricacy seems to exceed that in any of the other systems we have studied to date. This is doubtless due to the large basis table of E8 (at 2025 bases, a record) and its huge symmetry group (of over $10^{8}$ elements).\\

A comment should be made about the experimental measurements needed to realize the bases of E8, on which all the parity proofs of this paper depend. It might be asked if the projectors corresponding to some of the bases can be realized as simultaneous eigenstates of commuting three-qubit observables that are tensor products of Pauli operators of the individual qubits. While this is true of some of the bases, such as the ones we have identified as the Kernaghan-Peres sets, it is not true of the bases in general. The simplest way of generating an arbitrary basis from the computational basis is by following Eq.(\ref{eq1}) and applying a product of the appropriate powers of the three unitary operators $U,V$ and $W$. Designing efficient quantum gates for these operators is an interesting problem that we will not take up here. However it seems worth pursuing because a recent experiment\cite{canas2014b} has successfully generated several KS sets in a three-qubit system and holds out the possibility of eventually generating the more complex sorts of KS sets considered here.\\

It was pointed out in \cite{badziag} that any KS proof can be converted into an inequality that is satisfied by any noncontextual hidden variables theory but violated in measurements carried out on an arbitrary quantum state. It might be asked what the extent of the violation is for the parity proofs discussed in this paper. The answer to this question has already been given in an earlier work of ours\cite{Waegell2011b}. We showed there that for any basis-critical parity proof (i.e., one which fails if even a single basis is omitted from it), the upper bound of the inequality for any noncontextual hidden variable theory is $B-2$ (where $B$ is the number of bases in the proof) whereas quantum mechanics predicts the value of $B$.  This gap of 2 between the values predicted by hidden variable theories and quantum mechanics is a universal feature of all basis critical parity proofs. Thus the present proofs do not offer any particular advantage, from this point of view, over the many similar proofs\cite{Waegell2011c,Waegell2012a,Waegell2013,Waegell2011b,Waegell2014} we found earlier.\\

Gosset's polytope is the real representative of a complex polytope known as Witting's polytope\cite{leonardo}. Coxeter\cite{coxeter2} has carried out a systematic study of a large number of complex polytopes. It is possible that the ray systems derived from some of them might yield new proofs of the KS theorem. Whether this is true, and of what use it might be, are matters that remain to be explored.\\

{\bf Acknowledgements.} One of us (PKA) would like to thank David Richter for stimulating his interest in E8 and supplying him with a copy of Ref.\cite{richter}, which planted the seed for this work. \\


\end{document}